# Building Resilience to Climate Driven Extreme Events with Computing Innovations: A Convergence Accelerator Report

**December 2022**

## A Community Visioning Activity Organized By:

Elizabeth Bradley, University of Colorado - Boulder
Chandra Krintz, University of California - Santa Barbara
Melanie Moses, University of New Mexico

## With Support From:

Ann Schwartz, Computing Community Consortium
Aurali Dade, NSF Cognizant Program Director
Catherine Gill, Computing Community Consortium
Haley Griffin, Computing Community Consortium
Maddy Hunter, Computing Community Consortium

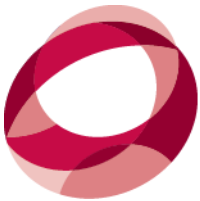 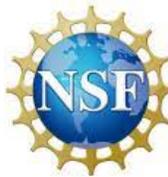

The material in this report is based upon work supported by the National Science Foundation under Grant No.1734706. Any opinions, findings, and conclusions or recommendations expressed in this material are those of the authors and do not necessarily reflect the views of the National Science Foundation.

# 1.  Introduction

In 2022, the National Science Foundation (NSF) funded the Computing Research Association (CRA) to conduct a workshop to frame and scope a potential Convergence Accelerator research track on the topic of "Building Resilience to Climate-Driven Extreme Events with Computing Innovations." The CRA's research visioning committee, the Computing Community Consortium (CCC), took on this task, organizing a two-part community workshop series, beginning with a small, in-person brainstorming meeting in Denver CO on 27-28 October 2022, followed by a virtual event on 10 November 2022. The overall objective was to develop ideas to facilitate convergence research on this critical topic and encourage collaboration among researchers across disciplines.

Based on the CCC community white paper entitled *Computing Research for the Climate Crisis*[1], we initially focused on five impact areas (i.e. application domains that are both important to society and critically affected by climate change):

- Energy
- Agriculture
- Environmental justice
- Transportation
- Physical infrastructure

We used these impact areas as a framework to help us identify participants in the workshop series and to plan our initial discussions and activities.

We used the workshops to discuss these impact areas and to collaboratively identify the necessary *building blocks* and key use-inspired *research thrusts* that can be brought to bear to address the complex challenges surrounding climate change. *Building blocks* are new abstractions, methods, and systems that can be used to facilitate and expedite technological innovation. *Research thrusts* are specific research directions identified by the participants as having potential for effecting positive change in a particular impact area. Research thrusts become building blocks if the participants identify them as being broad in technical scope and capable of being leveraged and specialized by a broad and diverse community of innovators to address seemingly disparate challenges *across* impact areas. Finally, our overarching goal with this effort was to identify computing research opportunities that can be developed and deployed following the timelines, guidelines, and goals described by the Convergence Accelerator program model.

We selected participants for the first workshop, in consultation with cognizant NSF program officers. The second, virtual workshop, was open to everyone. Both workshops included those with expertise in or across these impact areas. During the crafting of the participant lists—which appear in the appendices of this report—we also paid attention to demonstrated ability for interdisciplinary thinking, as well as to attaining a diverse and broad representation of the computing research community (demographics, institution type, and career stage).

---

[1] Bliss, N., Bradley, E., Monteleoni, C. (2021) *Computing Research for the Climate Crisis 2021*. https://cra.org/ccc/wp-content/uploads/sites/2/2021/08/Computing-Research-and-Climate-Change-%E2%80%94-August-2021.pdf

The goal of the in-person workshop was to refine the set of impact areas and identify research thrusts and building blocks. To enable this, we communicated with the participants in advance to establish the goals of the workshop: to brainstorm computational research that brings together collaborative multidisciplinary teams to create solutions with direct positive impact on climate change. In each impact area, we identified a "lead" (depicted in bold font in our participant list) who gave a brief presentation to define that area and frame some of the associated research challenges. The participants then went into breakout sessions to have focused discussions of research potential for that impact area, guided by three questions:

1. What are the key **building blocks** in computing research that are needed to expedite innovation in this impact area?
2. What use-inspired **research thrusts** can be brought to bear on this impact area to advance climate resilience?
3. What are the near-term **metrics** for success in this impact area?

Our intent with these questions was to build consensus around the climate crisis impact areas worthy of investment and building blocks that spanned areas (thus having potential for significant and potential near-term impact). We also asked the groups to construct an initial set of research thrusts for each area to indicate whether or not there is sufficient and necessary interest from the computing research community to pursue these thrusts as part of a Convergence Accelerator effort.

Following the breakout session, each group reported their findings to the whole group. These sequences (framing, breakout, report-back) were repeated for multiple impact areas. This was followed by a high-level synthesis discussion to review the topics and concepts that came up in multiple sessions, which culminated in a preliminary set of building blocks (listed below) and a concept matrix (depicted below) of research thrusts for the building blocks and the impact areas. Using this structure, we pursued a full-group brainstorming session in which participants used an electronic whiteboarding system (`mural.io`) to identify, discuss, and organize the research thrusts. This Mural represented the first major outcome of the brainstorming workshop.

**Computing research building blocks**
- ❖ Artificial intelligence (AI)
- ❖ Digital twins
- ❖ Cyberinfrastructure
- ❖ Optimization and planning
- ❖ User Interface/User Experience (UI/UX)
- ❖ Data

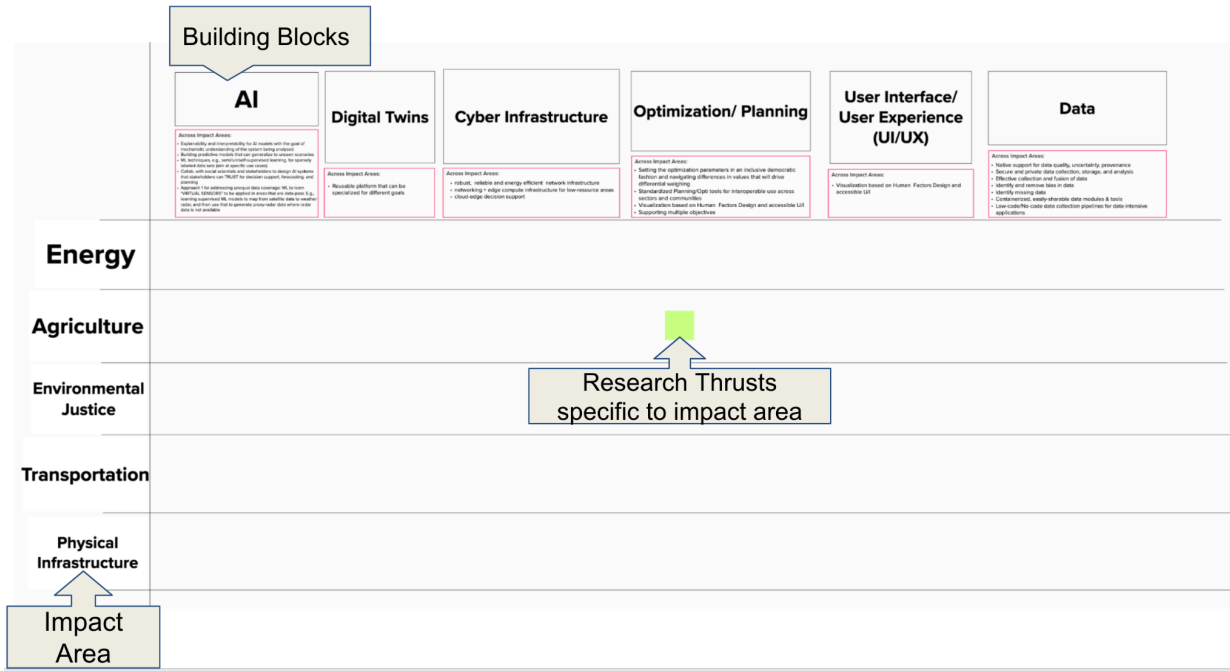

*Figure A-1. The skeleton of the concept matrix that was used in brainstorming during both workshops.*

A second, unplanned, outcome of this capstone discussion was a set of *cross-cutting principles*. This set emerged from the group discussions and brainstorming and consisted of principles that the group felt should be addressed by any research project in this area, regardless of impact area, building block, or research thrust.

We used these outcomes as the starting point for collaboration, discussion, and feedback at the second, community-wide, virtual workshop. To publicize this workshop, we released announcements on the CCC blog and Twitter account, the CRA Facebook and LinkedIn accounts, the Computing Research News, the NSF listserv, the Climate Informatics Google group, the ACM website, and the University of Colorado-Boulder and University of New Mexico websites. Almost 300 people registered for the workshop, of whom 122 attended.

The virtual workshop was held using Zoom two weeks after the in-person meeting. After a short overview from the organizers and the cognizant NSF program officers, we reviewed the outcomes from the in-person workshop and offered participants the opportunity to join a one-hour breakout session for the impact area of their choice, which were facilitated by one of the attendees from the in-person workshop. Using a shared Google document and a copy of the Mural row for that impact area, the participants in each breakout room began by working through the cross-cutting principles, offering suggestions as to what should be added, deleted, or changed. Each breakout group then moved to updating the Mural row with their suggestions about research thrusts. (If participants were unable to find an appropriate building block for their suggestions, we encouraged them to add new ones). After a short break, all participants returned to the main session and heard reports from each breakout room. This was followed by open discussion via the Zoom chat and Q&A channels and a wrap-up from the organizers. This second event produced revised and refined versions of the cross-cutting principles and the Mural.

The outcomes of this workshop series included a set of observations and recommendations that we document herein. We wish to note that the discussions at the meetings were very wide ranging. Given the timeline constraints of the Convergence Accelerator CFP, we distilled the community feedback and recommendations for those we thought were feasible, given these constraints. However, we want to emphasize that it became evident from these activities that the community feels strongly that significant and ongoing NSF investment is needed in this area to facilitate sustained progress, long-term impact, and societal benefits.

We next present these recommendations, together with supporting evidence from the workshop series. Slides for both the in-person and virtual events and the final concept matrix (in the form of a Mural) can be found in the appendix.

# 2. Recommendations for Building Resilience to Climate-Driven Extreme Events with Computing Innovations

We recommend that an NSF Convergence Accelerator (CA) track is warranted that focuses on computing advances that address the complex challenges surrounding climate change. Moreover, as part of the discussions and brainstorming in this community workshop series, we observed that there is tremendous opportunity for computing advances to be brought to bear on the climate crisis *across* impact areas. Both near-term and long-term computing research is needed to accomplish this. Since the Convergence Accelerator program focuses on the former, we identified key areas of overlap across impact areas and research thrusts to inform the recommendations in this report. Our recommendations include a set of:

- ❖ *Impact areas* – application domains that are both important to society and critically affected by climate change,
- ❖ *Research thrusts* with significant potential for addressing climate-induced challenges within and across impact areas,
- ❖ *Building blocks* – research advances that span impact areas and thus have potential for addressing multiple challenges concurrently, and
- ❖ *Cross-cutting principles* that all research projects should follow, regardless of impact area and technology.

## 2.1 Impact Areas

Our first recommendation is to integrate the *transportation* impact area into that of *physical infrastructure*, given the significant overlap in requirements, constraints, and climate impacts. Secondly, we recommend that environmental justice be a cross-cutting principle (described below) instead of an impact area, given the key role it plays across all of those areas.

Thus, the set of impact areas (our suggested application domain foci) that we recommend are:

- ❖ Energy – Energy system failures have economic impact as well as human costs. These harms are poised to worsen with the increased frequency and magnitude of extreme events due to climate change (hurricanes, heat waves, large rainfall events, etc.). At the same time, normal operating conditions for these systems are changing rapidly as renewables and other "edge" sources are added to the generation mix.
- ❖ Agriculture – Agriculture systems are sensitive to baseline shifts (e.g. changes in temperature and humidity, increased environmental variability, etc.) as well as to extreme events like droughts, floods, heat waves, and wildfires, which are projected to increase in both intensity and frequency as the climate changes.
- ❖ Physical infrastructure – Increased impacts of climate change and acceleration of frequency of extreme events stress regional power grids, transportation and communications networks, the manufacturing and financial services sectors and other aspects of the nation's critical infrastructure.

## 2.2 Research Thrusts and Building Blocks

The workshops enabled the research community to identify together a set of *building blocks* – areas of needed innovation that are key to combating the negative impacts of climate change across impact areas in the near term. The synthesized concept matrix that resulted from the workshops is depicted in this figure (a more readable version is available as an appendix). As described above, the columns in this matrix are the building blocks that the participants identified (AI, Digital Twins, Cyberinfrastructure, Optimization and Planning, UI/UX, and Data), and the rows are the climate crisis impact areas that we considered.

Figure A-2. *The final concept matrix, which combines all of the input from the in-person and virtual workshops.*

Within each cell (green box) in the concept matrix that was constructed with the Mural tool are the *research thrusts*—specific research directions identified by the participants as having potential for impact at the intersection of a particular impact area and a particular building block. The wide range of these thrusts clearly shows that our community sees many potential directions for addressing climate change impacts through the Convergence Accelerator program. Note that there are large clusters of research thrusts in some cells of the matrix. This is a direct reflection of opportunity and interest at these intersections.

Examples of research thrusts include:

❖ **Energy:**
  ➢ Using digital twins to make energy infrastructure more robust and resilient
  ➢ Utilizing a geographically-aware budget to reduce peak demand, allowing the minimization of rolling blackouts during heatwaves and cold snaps
  ➢ Planning and optimization based on input from affected communities
❖ **Agriculture:**
  ➢ AI tools for pest and invasive species detection
  ➢ Edge computing on mobile platforms for labor shortage mitigation
  ➢ Optimization advances for tailoring agriculture to climate change (e.g. what should crops look like in 2035, so that they are resilient to extreme events?)
  ➢ Simple, intuitive UIs for precision agriculture tools which don't require the user to be an expert in computer or agricultural sciences.
❖ **Physical Infrastructure:**
  ➢ A standardized platform for the recovery process for all disasters (FEMA, HUD CDBG, etc.)

- ➢ Real-time health monitoring for optimal repair schedules
- ➢ Cyber-physical systems: e.g., active structures that adapt stiffness for resilience to threats, such as high wind, etc.
- ➢ Resource efficient and privacy preserving analytics
- ➢ Eco-friendly decision support for travelers (e.g. directing a user to take a route that adds 9 minutes to their journey, but reduces their fuel consumption by 25%)
- ➢ Using data visualization as an educational tool to change behaviors

The PI team then performed extensive analysis of these research thrusts to identify those that occurred repeatedly across impact areas, and therefore are the major points of opportunity for near-term computing research to support climate adaptation. These building blocks, which are separated out into lists below each column header in Figure 1, are listed below:

AI

- Explainability and interpretability for AI models with the goal of mechanistic understanding of the system being analyzed
- Predictive models that can generalize to unseen scenarios
- Hybrid AI that combines physics-based with data-driven models
- Techniques that address spatially unequal data coverage/availability
- Physics-guided AI to generate and integrate theoretical & empirical data to improve modeling, analysis, and explainability
- Consideration of randomness, unmeasured, and unexpected factors
- Manage trade-offs to permit ML/AI at the edge/in-situ
- AI to design sustainable and resilient materials
- Trustworthy and robust AI

Digital Twins

- Clear definition of Digital Twins' purpose, goals, and validation
- Reusable cloud+edge platform that can be specialized for different goals
- Support for exploration of what-if scenarios
- Advances for interoperability (devices, models, services)
- Couple GIS + simulation into platform
- Probabilistic programming to build rich simulators & reason about possible trajectories
- Integrated data fusion from multiple sources, modalities, and scales
- Support for performance tuning and accuracy trade-offs

Cyberinfrastructure

- Robust, reliable, and energy efficient network infrastructure
- Networking + edge compute infrastructure for low-resource areas
- Cloud-edge decision support

- Service-ization (modularity) to enable maintainability/sustainability
- Support for metric collection & trade off management (power, size, performance)
- Biodegradable materials (or non-rare-earths) for computing devices and components

Optimization/Planning

- Setting the optimization parameters in an inclusive democratic fashion and navigating differences in values that drive differential weighting
- Standardized tools for interoperability across sectors & communities
- Support for multiple objectives
- Support for interactive visualization of decision support data and processes

UI/UX

- Visualization based on human factors design and accessible U/I
- Specialized for multiple, heterogeneous affected community targets & needs
- Decision-support tools that present a range of options and tradeoffs in an understandable way
- Psychology-grounded tools to help people think about risks and costs
- Tools to help people share information and tasks

Data

- Support for managing data quality, uncertainty, provenance
- Secure & private data collection, storage, protection, ownership management, and analysis
- Effective & efficient collection, storage compression, and fusion of data from multiple sources, modalities, & scales
- Identify and remove bias in data
- Identify missing data
- Containerized, easily-shareable, low-code/no-code data pipelines, modules & tools
- Data assimilation from sensors and/or crowd sourcing for rapid updating
- Statistical Design for responsive data collection and synthetic data generation
- Support for open knowledge networks

As described in the Convergence Accelerator program description, these building blocks represent research that is broad in technical scope, has far-reaching impact on society, builds upon foundational research, and requires a multidisciplinary, convergent research approach to be successful. Moreover, because they span impact areas, these building blocks have significant potential for affecting multiple use cases of societal import if pursued and invested in.

## 2.3 Cross-Cutting Principles

As mentioned in the first section of this report, the participants in this visioning activity identified an important set of cross cutting principles that Convergence Accelerators should address. For any research thrust or set of building blocks, projects should employ a systems-level approach, involving affected communities and addressing environmental justice. Projects should also develop an actionable set of outcomes and metrics. Details appear below.

- *Employ a systems-level approach that considers resilience, usability, trustworthiness, and explainability in the design.*

  The human/climate system is complicated, nonlinear, nonstationary, and highly coupled. Computational solutions need to respect this; addressing elements of the associated problems in isolation will not work because of uncertainties and couplings, both known and unknown, that may have unintended consequences. Finally, there is a need to consider socio-technical issues across all phases of the research. As mentioned in the recent National Academies report on responsible computing[2], "It is much easier to design a technology correctly from the start than it is to fix it later." This will require meaningful, operationalized partnerships between computer scientists on the project teams and experts in the social and behavioral sciences.

- *Include a plan for identifying and involving a wide range of affected communities, across all phases of the project, including design, deployment, and adoption.*

  In order to assure that the solutions are a deep match to all aspects of the associated problems, affected communities must be involved in meaningful ways in all phases of projects funded under this program, including those historically underserved. This will create many benefits: not only building support and facilitating adoption, but also avoiding data abuse/misuse issues and creating tools that the target audiences trust. A challenge here will be identifying those target audiences—e.g., end users, developers/innovators, students—and working with them to understand the associated usability issues. Ideally, there should be incentives and funding mechanisms in place to foster sustained engagement and continuous improvement: e.g., ongoing feedback loops between the researchers and the community members, all the way out to the field workers.  This will, of course, be a tall order, given the constraints of the Convergence Accelerator program, but research funded under this program could provide effective nucleation points for the types of follow-on activities and sustained support and investment that will be necessary for real impact on these novel, difficult, and critical problems.

---

[2] *Fostering Responsible Computing Research*. National Academies Press, Oct. 2022. *Crossref*, https://doi.org/10.17226/26507.

- *Provide a set of actionable outcomes for the project*

  These should include goals, milestones, and deliverables. Solutions must also be able to address a societal need and be capable of growing and adapting to achieve this in the long term.

- *Include well-defined metrics for success of the project that take into account not only **direct** climate impacts, but also an appropriate subset of the following considerations:*

  - *The impact of computing itself on the climate*

    Computation not only has a significant carbon footprint, but also major supply-chain and e-waste issues. Training a modern machine-learning model, for instance, can generate as much CO2 as five cars will produce over their lifetimes[3]. Mining of the rare earths used in computer electronics can be environmentally detrimental and disposing of that equipment properly is a real challenge. These factors must be identified and considered during all research funded by this program, across the full stack: hardware, software, cyberinfrastructure (e.g., the cloud), etc.

  - *Human impacts*

    Solutions to climate-change issues cannot have impact if they are not adopted, used, understood, and trusted by their target audiences. This requires effort to be devoted (again, throughout the design and deployment phases) to human-centered design thinking: satisfaction, inclusivity, cultural relevance.

  - *Tech transfer*

    Uptake by industry and spinoff of startups, perhaps catalyzed by academic-industry partnerships, could greatly leverage the impact of climate-change solutions developed under the auspices of this program.

  - *The complicated, often-conflicting nature of the goals that can arise in this assessment*

    A successful systems-level approach to climate-change problems will require multi-objective planning and optimization that takes into account risk and uncertainty, as well as differences in metrics across different affected and targeted groups, to give economical, environmental, and societal resiliences appropriate weight in decisions.

---

[3] https://www.forbes.com/sites/robtoews/2020/06/17/deep-learnings-climate-change-problem

- *Consider and actively promote environmental justice*

    - Any and all proposed technological advances to address climate change must simultaneously work to address, and ultimately overturn, unequal environmental legacies, as well as to assure equity going forward.

Given the constraints of the Convergence Accelerator program, it will not be practical for every project to address every element of these cross-cutting principles in any comprehensive way, but all proposals should identify the critical subset of these principles that will guide their project, and **all** projects should include a meaningful set of metrics for their work.

## 2.4 Recommendations

In summary, given our experience with this workshop series, we recommend that NSF create a Convergence Accelerator track in support of computing advances that facilitate adaptation and build resilience to climate change. In collaboration with the research community, we have identified a set of cross-cutting principles that we believe all Convergence Accelerators should address. We have also identified key impact areas on which use-inspired research can focus, as well as computing research building blocks that build upon foundational research and require a multidisciplinary, convergent research approach in order to succeed in producing positive and far-reaching impacts for society. Finally, we recommend significant and ongoing NSF investment in climate-focused computing research in order to facilitate sustained progress, long-term impact, and societal benefits.

# Appendix A. Workshop Materials

## Figure A-1. Blank Mural

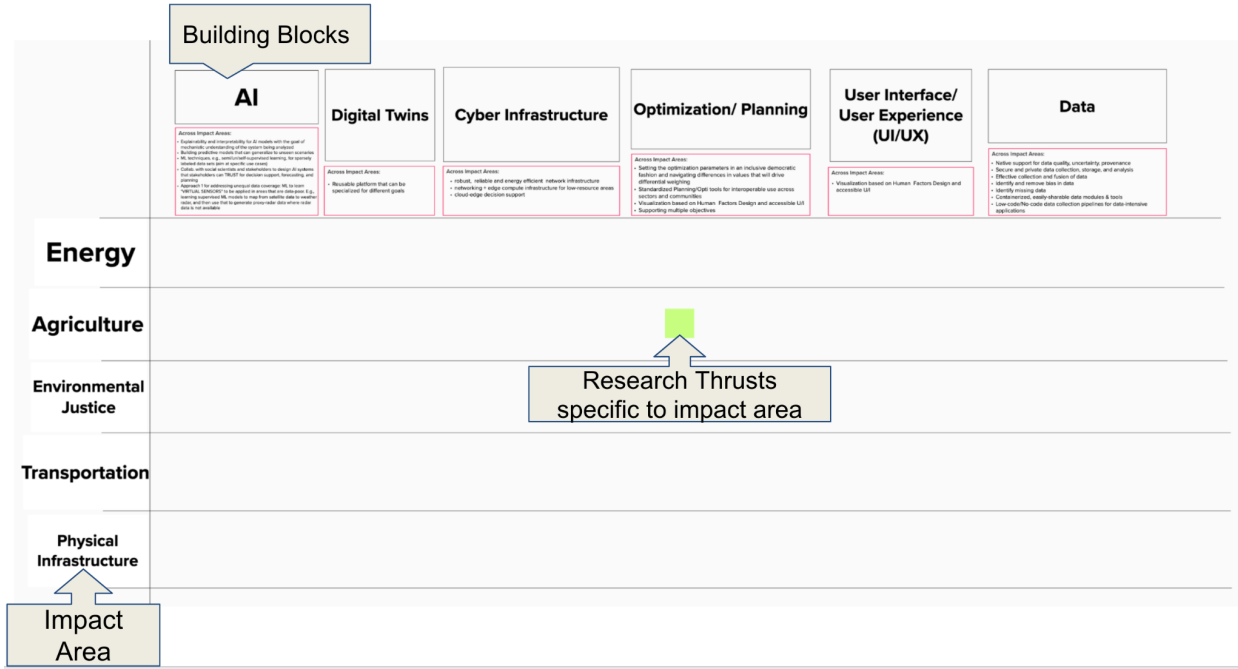

Figure A-1. A blank copy of the concept matrix that was used in brainstorming during both workshops. This copy has labels indicating the names we have chosen for the different categories in the matrix.

## Figure A-2. Final Concept Matrix

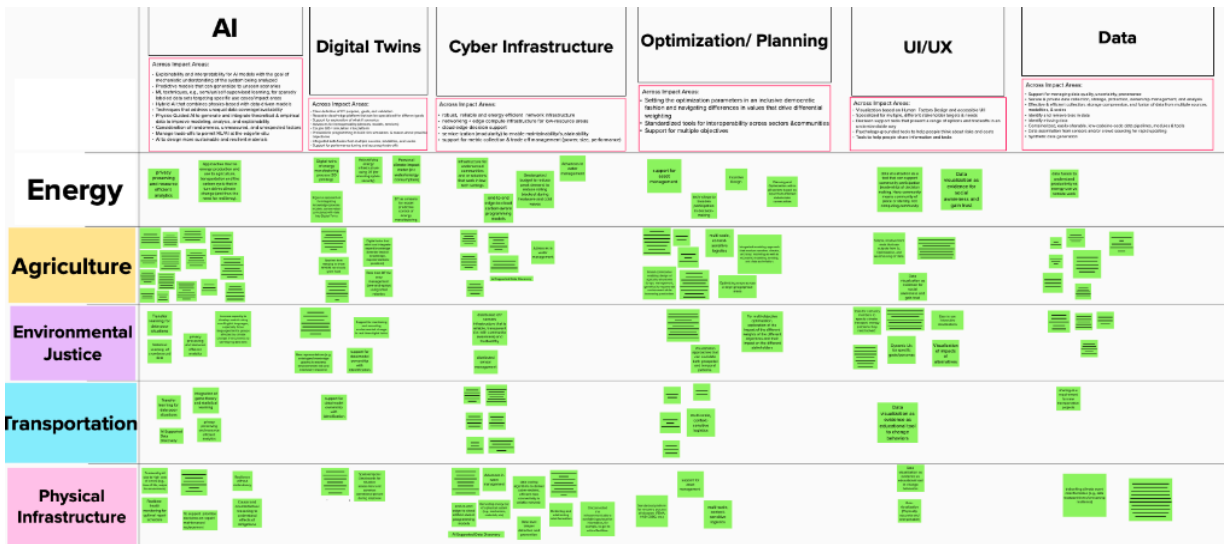

Figure A-2. The final concept matrix, which combines all of the input from the in-person and virtual workshops.

**Figure A-3. In-Person Workshop Slides**

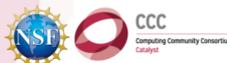

**Building Resilience to Climate Driven Extreme Events with Computing Innovations: A Convergence Accelerator Workshop**

*October 27-28, 2022*

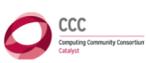

## Agenda Day 1

October 27, 2022 (Thursday)

| | |
|---|---|
| 12:00 PM | **Lunch Available** \| Lodo |
| 01:00 PM | **Welcome and Introductions** \| Central City |
| 01:30 PM | **Topic 1: Energy** \| Central City |
| 02:30 PM | **Topic 2: Agriculture** \| Central City |
| 03:30 PM | **BREAK** \| Central City |
| 04:00 PM | **Topic 3: Transportation** \| Central City |
| 05:00 PM | **General/Cross-Cutting Discussion** \| Central City |
| 05:30 PM | **Wrap up** \| Central City |
| 06:30 PM | **Dinner** \| Lodo |

## Agenda Day 2

October 28, 2022 (Friday)

| | |
|---|---|
| 07:30 AM | **Breakfast** \| Lodo Room |
| 08:30 AM | **Recap of Day 1** \| Central City Room |
| 08:45 AM | **Discussion: Building Blocks** \| Central City Room |
| 09:15 AM | **Panel 4: Environmental Justice** \| Central City Room. Breakouts in Leadville and Silverplume. |
| 10:15 AM | **BREAK** \| Central City Room |
| 10:30 AM | **Concluding Comments/Discussion** \| Central City Room |
| 10:45 AM | **What have we not discussed?** \| Central City Room |
| 11:15 AM | **Deliverable Muraling** \| Central City Room |
| 12:00 PM | **Lunch** \| Lodo Room |
| 01:00 PM | **End of Workshop** |

## Welcome Remarks

**Liz Bradley, University of Colorado-Boulder**
**Chandra Krintz, UC Santa Barbara/AppScale Systems Inc.**

## Lightning Introductions

## Organizers

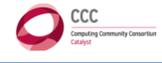
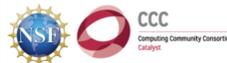
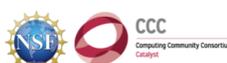
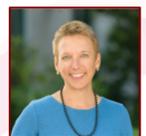

**Chandra Krintz**
University of California
Santa Barbara

**Liz Bradley**
University of
Colorado-Boulder

**Melanie Moses**
University of New Mexico

**Aurali Dade**
NSF (TIP/ITE)

## With Support From

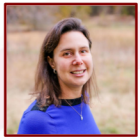
**Ann Schwartz**
CCC

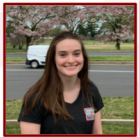
**Catherine Gill**
CCC

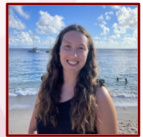
**Haley Griffin**
CCC

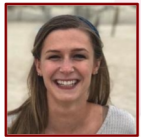
**Maddy Hunter**
CCC

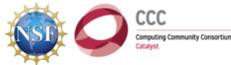

## Welcome from the NSF

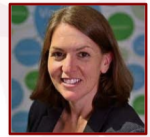

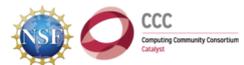

---

NSF's Convergence Accelerator

**NSF'S CONVERGENCE ACCELERATOR**

# 2022 WORKSHOP POINTS OF EMPHASIS

Tuesday October 27, 2022

Aurali Dade, Program Director
Convergence Accelerator
Directorate for Technology, Innovation and Partnerships
National Science Foundation

ACCELERATING CONVERGENT SOLUTIONS FOR SOCIETAL IMPACT

---

## A Pivotal Moment for the Nation

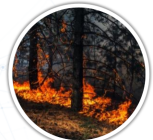
**Climate change**

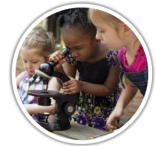
**Equitable access to education, health care**

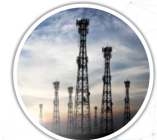
**Critical and resilient infrastructure**

---

## A Pivotal Moment for Science & Engineering

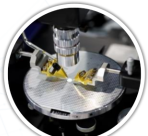
**Pace of discovery accelerated by data, emerging technologies**

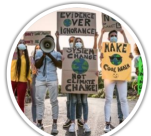
**Demand for societal impact**

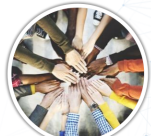
**Opportunity to leverage partnerships**

---

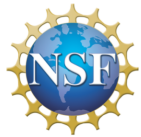

**MISSION:**
To promote the progress of science; to advance the national health, prosperity, and welfare; and to secure the national defense; and for other purposes

**VISION:**
Envisions a nation that capitalizes on new concepts in science and engineering and provides global leadership in advancing research and education

---

# TIP Directorate for Technology, Innovation and Partnerships
Creates breakthrough technologies | Meets national needs | Empowers all

---

# CONVERGENCE RESEARCH
Today's grand challenges will **NOT** be solved by one discipline working alone.

Grand Challenges require
**CONVERGENCE:**
the merging of ideas, approaches, and technologies from widely diverse fields of knowledge to stimulate innovation and discovery.

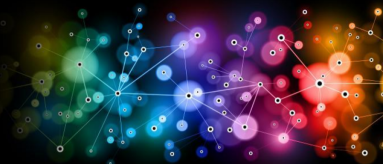

# NSF Convergence Accelerator
*Accelerating Solutions Toward Societal Impact*

**MISSION:**
The Convergence Accelerator speeds the transition of convergence research into practice to address national-scale societal challenges

**VISION:**
Convergence research and multi-institutional teams that include users and other stakeholders will provide high-impact results to societies at scale

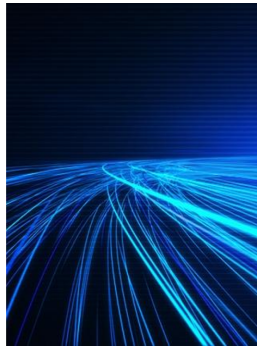

# NSF Convergence Accelerator
*Accelerating Solutions Toward Societal Impact*

**GOALS:**
- Disrupt the usual way of NSF business through a new innovation model
- Expand and diversifies multidisciplinary teams and partnerships to include academia, industry, non-profits, government, and other sectors
- Deliver solutions that have a national societal impact

| Characteristics | Proactively & Intentionally Managed |
|---|---|
| Use-inspired research | Teams and Cohorts—"Tracks" |
| Clear goals, milestones, high-impact deliverables | Cooperation and Competition |
| Leverages multidisciplinary teams | Intensive education and mentorship—human-centered design thinking, team science, and customer discovery |
| Larger, national societal scale | Mission-driven evaluation |
| Requires diverse partnerships – industry, non-profits, academia | |
| Acceleration at speed and scale | |

## CONVERGENCE ACCELERATOR PROGRAM

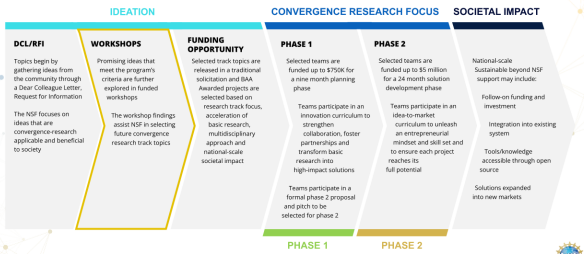

## CONVERGENCE ACCELERATOR EXAMPLE
*2021 Cohort: Ideation to Convergence Research*

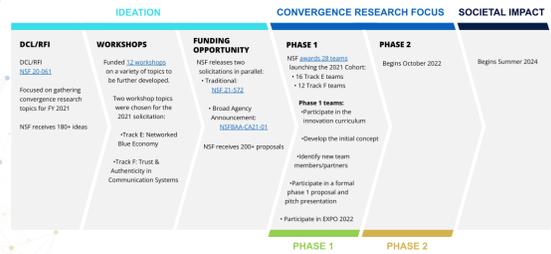

## PROGRAM STRUCTURE

**IDEATION (DCL/RFI, WORKSHOPS):**
Selected by gathering input from the community. Identified topics must meet a societal need at scale, be built upon foundational research, and be suitable for a multidisciplinary, convergence research approach.

**PHASE 1 (PLANNING):**
Up to $750K over 9 months is provided to further develop the initial concept (building upon basic research), identify new team members/partners, participate in a hands-on innovation curriculum, and develop an initial/low-fidelity prototype.

**PHASE 2 (IMPLEMENTATION):**
Up to $5M over 24 months to develop solution prototypes and to build a sustainability model to continue impact beyond NSF support.

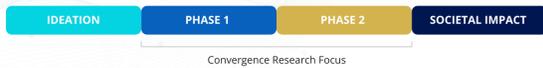

Convergence Research Focus

## Convergence Accelerator Portfolio

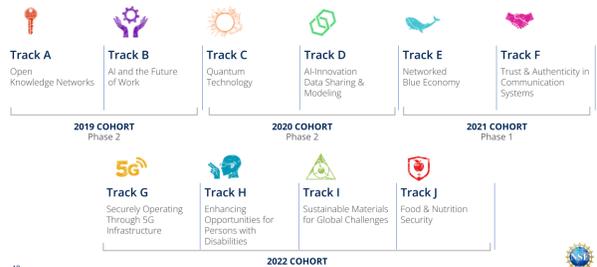

## 2023 Cohort Track Topics – TBD

If selected as a topic for the 2023 solicitation, this workshop report will be used as a reference for the community to make sure it's as clear, concise, and comprehensive as possible.

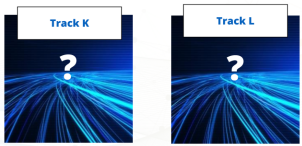

## Convergence Accelerator
## Workshop Charge–Points of Emphasis

*As you engage in the workshop discussion, keep in mind the following aspects of the Convergence Accelerator:*

**Convergence:** Multiple disciplines with a focus on social science aspects; think big—don't just include experts from a single institution or discipline

**Cross-cutting Partnerships:** Multiple disciplines, organizations, and sectors; not just academia; must include industry, non-profits, government, and other communities of practice
- Diverse partnerships provide valuable expertise and insights to position the deliverable for success.
- Multidisciplinary approach (different sectors and expertise)
- Use-inspired research (end-user and prototyping research)

**Broadening Participation:** Must include the participation of underrepresented groups (e.g., expertise, partnerships, user groups, resource needs)

## Convergence Accelerator
## Workshop Charge–Points of Emphasis Continued

*As you engage in the workshop discussion, keep in mind the following aspects of the Convergence Accelerator:*

**Deliverables:** What can be delivered to the American people in 3 years, (e.g., Prototypes); What impact will the solutions have on a national scale?

- Research papers are **not** sufficient.
- Deliverables do not have to result in commercialization, but they must be useful and needed tools, test beds, living labs, etc.

**Track Alignment:**

- How can multiple funded teams work together to solve a national-scale complex challenge?
- Each track funds a set of diverse teams focusing on different aspects of a national-scale societal challenge
- Teams are uniquely positioned to ensure the highest societal impact

---

## CONVERGENCE ACCELERATOR PROGRAM TIMELINE
Overall Timeline – 2019–2023

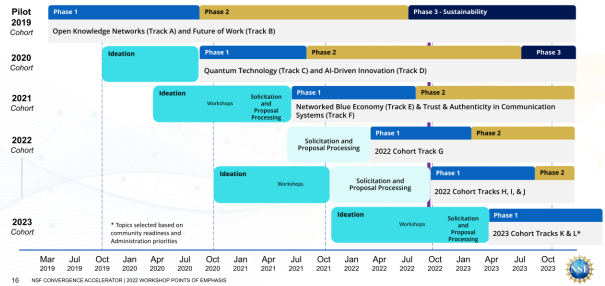

---

## Workshop Themes for 2023 Track Topics

---

## Workshop Themes for 2023 Track Topics

---

## NSF Convergence Accelerator Team

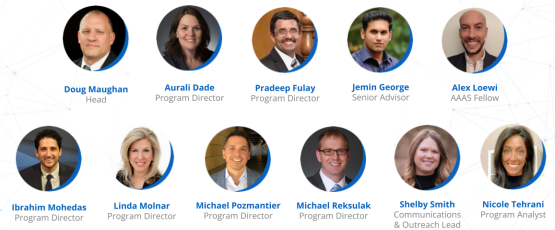

---

**LEARN ABOUT US.**

- Unique Program Model
- Exciting Program Offerings
- Current Portfolio
- Ways to Connect into the Program and with the Team

**BE THE FIRST TO KNOW.**
Join our mailing list to learn about important program updates and new funding opportunities!

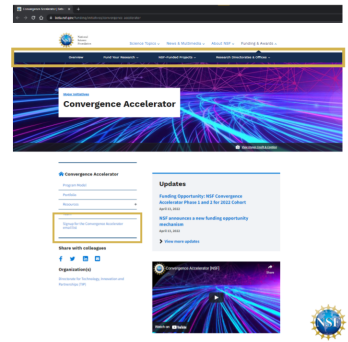

---

# THANK YOU

beta.nsf.gov/funding/initiatives/convergence-accelerator

Convergence-Accelerator@nsf.gov

**Aurali Dade, Program Director**
NSF Convergence Accelerator
adade@nsf.gov

---

## TIP: Accelerating Research Toward Impact

| Fostering Innovation and Technology Ecosystems | Establishing Translation Pathways | Partnering to Engage the Nation's Diverse Talent |
|---|---|---|
| Nurtures regional and national innovation and technology ecosystems to support researchers and innovators to converge, develop and accelerate use-inspired research for societal impact. | Supports startups through a lab-to-market platform and establishes new pathways for translating research results for society. | Advances and deepens high-impact, public and private partnerships across all areas of science, engineering and education to cultivate innovation ecosystems, create technology solutions, and support future STEM leaders. |

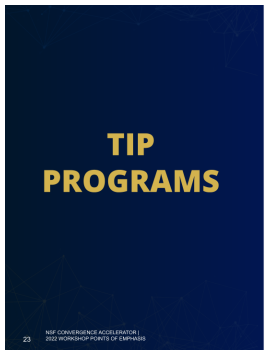

**TIP PROGRAMS**

- America's Seed Fund powered by NSF (SBIR/STTR)
- Convergence Accelerator
- Innovation Corps (I-Corps™)
- Partnerships for Innovation (PFI)
- Pathways to Enable Open-Source Ecosystems (POSE)
- Regional Innovation Engines (NSF Engines)

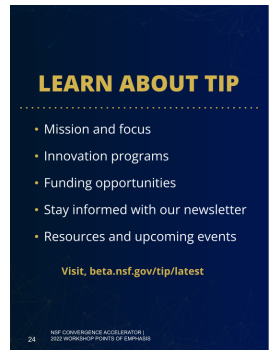

**LEARN ABOUT TIP**

- Mission and focus
- Innovation programs
- Funding opportunities
- Stay informed with our newsletter
- Resources and upcoming events

Visit, beta.nsf.gov/tip/latest

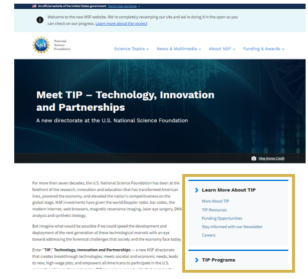

**Meet TIP – Technology, Innovation and Partnerships**

---

## COMPUTING COMMUNITY CONSORTIUM

The mission of Computing Research Association's Computing Community Consortium (CCC) is to enable the pursuit of innovative, high-impact computing research that aligns with pressing national and global challenges.

**Who**
- Council - 23 members
- CCC/CRA Staff
- Chair, VC, & Director

Daniel Lopresti
CCC Chair

**Inputs: Bottom-up, Internal, & Top-Down**

**What:**
- Workshops & Conf. Blue Sky Tracks
- Whitepapers & Social Media
- Reports Out (esp. to government)
- Symposium for DC'ers

Nadya Bliss
CCC Vice Chair

**Human Development**
- Early Career Workshops & Participation
- Council Membership
- Leadership w/ Gov't (LISPI)

Elizabeth Bradley
CCC Chair Emerita

---

## CCC VISIONING WORKSHOPS...

- Engage the community, together with relevant stakeholders, rapidly capturing and synthesizing the important ideas
- Facilitate broad thinking with compelling examples
- Create new avenues for (interdisciplinary) collaboration
- Frame future opportunities in a manner that energizes the community and engages potential funders
- Align with national and computing research priorities
- Articulate needs and barriers to research impact

---

## Goals & Expected Outcomes

- Develop and frame ideas to incorporate convergence research and encourage collaboration among stakeholders across disciplines/experiences
- Identify research tracks to include in next year's program solicitation
  - Research Focus: *Building Resilience to Climate Driven Extreme Events with Computing Innovations*

- This workshop: draft an outline for this focus and its tracks
  - Evolve/Improve with wider community at **virtual meetup on Nov 10**

- Identify key **computing research building blocks** that span impact areas
  - Energy, agriculture, transportation, environmental justice, infrastructure, and more…
  - To expedite innovation and near term demonstrable impact

---

## Workshop Structure

- Four 1-hour sessions (5 + 35 + 5 + 15)
  - Area framed briefly by expert in the field (5mins)
    - Today: Energy (1:30), Agriculture (2:30), Transportation (4pm)
    - Tomorrow: Environmental Justice
  - Breakout rooms (35mins): Brainstorm computing advances (existing and new)
    - **Use inspired solutions** vs technologies
    - End-to-end, solutions themselves must be climate resilient
    - CCC staff will take shared notes
  - Regroup: Reporters summarize discussion from each breakout (15mins)

- Active participation in breakouts outside of your area of expertise
  - Get outside of your comfort zone

---

Figure A-3. In-Person Workshop Slides.

**Figure A-4. Virtual Workshop Slides**

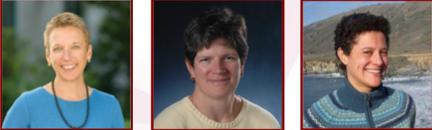

## Building Resilience to Climate Driven Extreme Events with Computing Innovations: A Virtual Convergence Accelerator Workshop

**12:00-3:00 pm EST**

### CCC Welcome

Chandra Krintz
University of California
Santa Barbara

Liz Bradley
University of
Colorado-Boulder

Melanie Moses
University of New Mexico

Ann Schwartz

Catherine Gill

Haley Griffin

Maddy Hunter

### COMPUTING COMMUNITY CONSORTIUM

The mission of Computing Research Association's Computing Community Consortium (CCC) is to enable the pursuit of innovative, high-impact computing research that aligns with pressing national and global challenges.

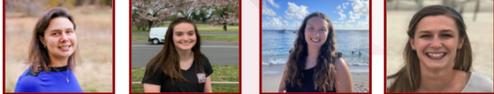

**Who**
- Council - 23 members
- CCC/CRA Staff
- Chair, VC, & Director

**Inputs: Bottom-up, Internal, & Top-Down**

**What:**
- Workshops & Conf. Blue Sky Tracks
- Whitepapers & Social Media
- Reports Out (esp. to government)
- Symposium for DC'ers

**Human Development**
- Early Career Workshops & Participation
- Council Membership
- Leadership w/ Gov't (LISPI)

Daniel Lopresti
CCC Chair

Nadya Bliss
CCC Vice Chair

Elizabeth Bradley
CCC Chair Emerita

### NSF Welcome and Overview

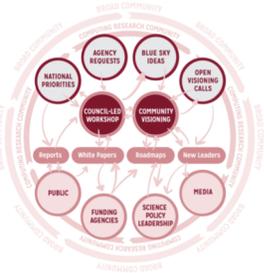

Aurali Dade
The National Science
Foundation
TIP/ITE

### Goals, Process, & Expected Outcomes

- Develop and frame ideas to incorporate convergence research and encourage collaboration among stakeholders across disciplines/experiences

- Identify research tracks for potential inclusion in next year's program solicitation
  - Research Focus: *Building Resilience to Climate Driven Extreme Events with Computing Innovations*

- Oct 27 "pre-workshop": Small in-person event to frame research focus
  - **Impact areas** (application domains)
  - Computing research **building blocks** that span impact areas
  - **Research thrusts** within impact areas
  - **Cross-cutting principles**

- **Today's event:** Brainstorm and build upon framing to draft NSF report

### Pre-Workshop: When & Who

**Participants**
- Vikram Adve
- Sujata Banerjee
- David Begay
- Tracy Camp
- Michael Dunaway
- Tayo Fabusuyi
- Baskar Ganapathysubramanian
- Peter Harsha
- Raya Horesh
- Daniel Jacobson
- David Jensen
- Vipin Kumar
- Christine Lv
- Claudia Marin
- Charlie Messina
- Aditi Misra
- Claire Monteleoni
- Raj Pandya
- Shashi Shekhar
- Jiayang Sun

27-28 October 2022
Denver CO

## Pre-Workshop: Impact Areas

- Energy
- Agriculture
- Transportation
- Environmental Justice
- Physical Infrastructure
- *Your impact area(s) here…*

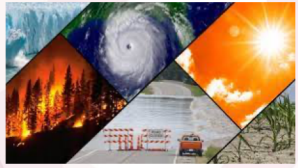
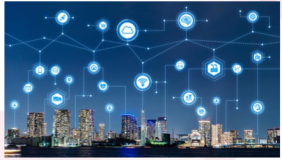
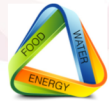
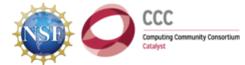

## Pre-Workshop: Building Blocks

- Six broad computing research **building blocks across those impact areas**:

  ➢ AI
  ➢ Digital Twins
  ➢ Cyberinfrastructure
  ➢ Optimization/Planning
  ➢ Visualization
  ➢ Data
  ➢ …?

## Mural

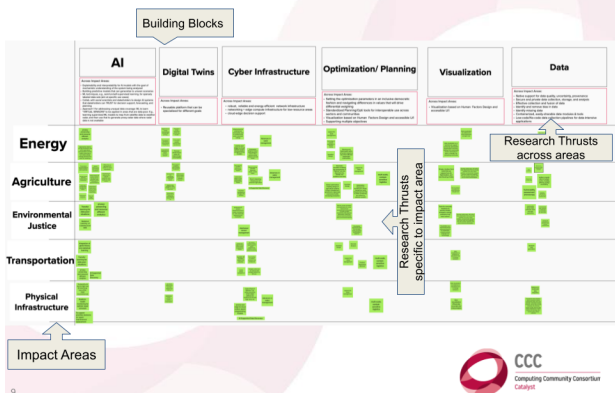
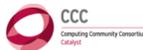
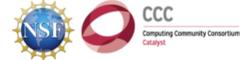

## Pre-Workshop: Building Blocks – Research Thrusts Across Impact Areas

**AI**
- explainable, interpretable, use inspired
- generalizable predictive models
- semi/un/supervised ML + sparse labeling
- stakeholder involvement to build trust
- transferable from data rich to data poor

**Digital Twins**
- uniform, reusable platform
- customizable for different goals

**Cyberinfrastructure**
- robust + easy of use and extend
- resource constrained, energy efficient
- edge-aware network, storage, compute

**Optimization/Planning**
- democratize parameterization & weighting
- standardization
- multi-objective support

**Visualization**
- human factors design & accessibility

**Data**
- quality, uncertainty, provenance
- sharing, security & privacy
- collection, fusion & bias removal
- tools & pipelines

## Pre-Workshop: Cross-Cutting Principles

- Considers the impact of computing itself on the climate.

- Includes well defined metrics for success

- Employs a holistic, end-to-end systems-level approach that incorporates resilience considerations

- Addresses usability for different user targets

- Includes a plan for stakeholder involvement *and participation* (including those historically underserved)

- Integrated communication, outreach, and adoption plan

- …?

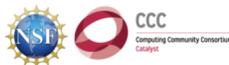

## Agenda

- 12:00 - General Overview
  - CCC Welcome
  - NSF Welcome and overview of Convergence Accelerator Program
  - Overview of Part 1 of Workshop
  - Charge for Breakout Discussions
- 12:30 - Topical Breakouts
  - Energy, Agriculture, Environmental Justice, Transportation, Infrastructure
- 1:30 - Break
- 1:45 - General Session
  - Reports from Breakout Discussions
  - Wrap up

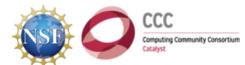

### Instructions for Breakout Rooms

- Zoom room per impact area (1-2 breakouts created there on-demand)
- Identify "Reporter"
- Document discussion/ideas via shared google doc & mural row
- Discuss cross-cutting principles (15mins)
- Mural work: brainstorm research thrusts (30mins)
  - Select *Enter as a visitor*
- Refine report-out plan and identify what is missing (15mins)
- Break

Report-outs & discussion starts at 1:45pm ET at original webinar link

### Please join a breakout room! We will meet back here at 1:45 pm for reports and discussion

- Agriculture room: cra.org/ccc/Agriculture

- Energy room: cra.org/ccc/Energy

- Transportation room: cra.org/ccc/Transportation

- Environmental Justice room: cra.org/ccc/Environmental-Justice

- Infrastructure room: cra.org/ccc/Infrastructure

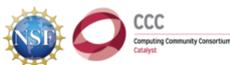
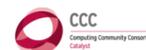

### Break

- 15 minute break; please be back by 1:45 pm EST to hear reports from each breakout room!

- Agriculture room: cra.org/ccc/Agriculture
- Energy room: cra.org/ccc/Energy
- Transportation room: cra.org/ccc/Transportation
- Environmental Justice room: cra.org/ccc/Environmental-Justice
- Infrastructure room: cra.org/ccc/Infrastructure

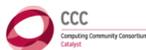

### Report out and discussion

- Three minutes per breakout room

- Grouped by impact area: first report(s), then discussion
  - Energy, Agriculture, Environmental Justice, Transportation, Infrastructure
  - Use zoom's chat feature for comments, questions, responses

- You can also put your comments and questions in the google doc from your breakout room

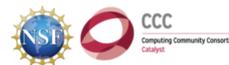

### Wrap-up

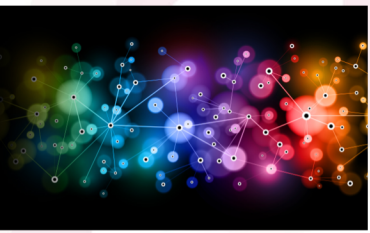

Grand Challenges require **CONVERGENCE:** the merging of ideas, approaches, and technologies from widely diverse fields of knowledge to stimulate innovation and discovery.

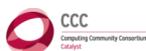

### THANK YOU!

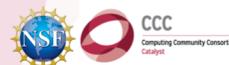

Figure A-4. Virtual Workshop Slides

# Appendix B. Participant Information

**Figure B-1. Virtual Workshop Registrants by Sector**

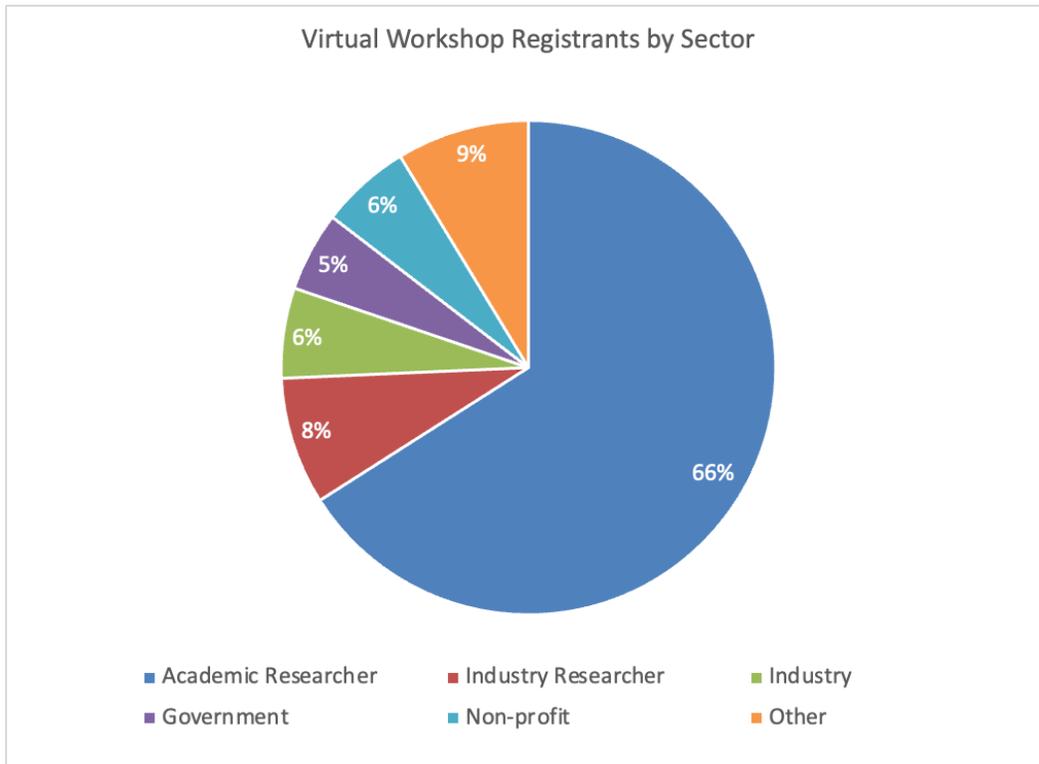

Figure B-1. A graph of the virtual workshop registrants and the sector in which they work.

**Figure B-2. Virtual Workshop Registrants' First Choice of Topic**

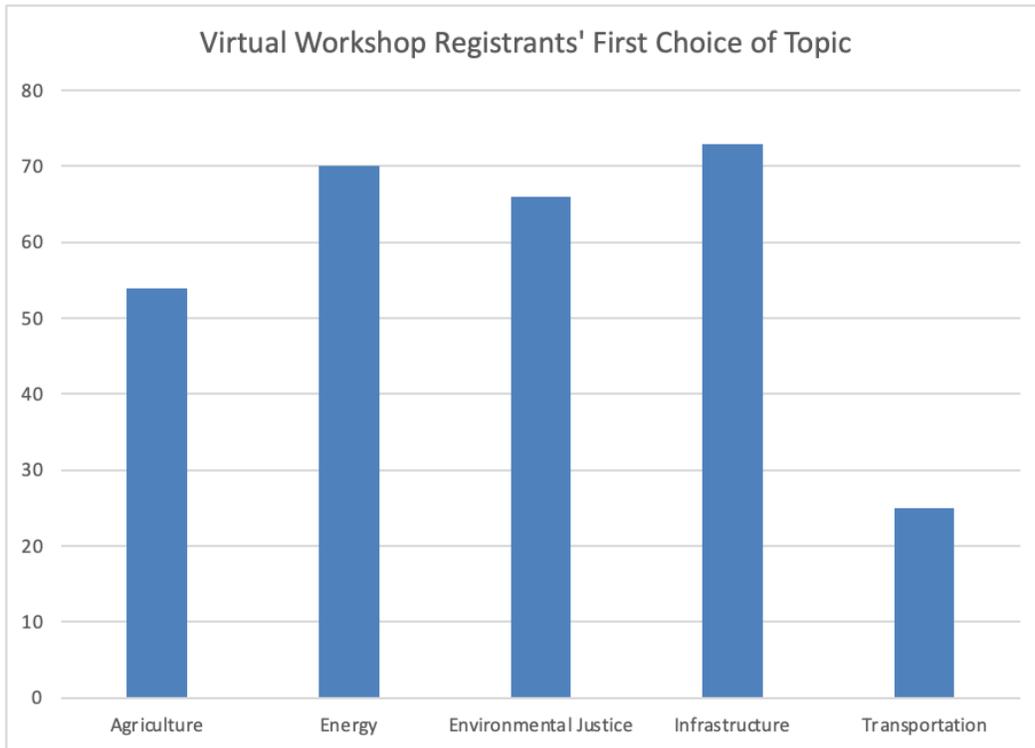

Figure B-2. A graph of the virtual workshop registrants' first choices of topics to discuss during the workshop.

**Figure B-3. List of In-Person Workshop Attendees**

| In-Person Workshop Attendees | | |
|---|---|---|
| **First Name** | **Last Name** | **Institution** |
| **Vikram** | **Adve** | University of Illinois at Urbana-Champaign |
| Sujata | Banerjee | VMware |
| David | Begay | University of New Mexico |
| Elizabeth | Bradley | University of Colorado - Boulder |
| Tracy | Camp | Computing Research Association |
| Aurali | Dade | National Science Foundation (ITE/TIP) |
| Michael | Dunaway | National Institute of Standards and Technology |
| **Tayo** | **Fabusuyi** | University of Michigan |
| Baskar | Ganapathysubramanian | Iowa State University |
| Catherine | Gill | Computing Community Consortium |
| Haley | Griffin | Computing Community Consortium |
| Peter | Harsha | Computing Research Association |
| **Raya** | **Horesh** | IBM Research |
| **Daniel** | **Jacobson** | Oak Ridge National Laboratory |
| David | Jensen | University of Massachusetts - Amherst |
| Chandra | Krintz | University of California - Santa Barbara |
| Vipin | Kumar | University of Minnesota |
| Christine | Qin | University of Colorado - Boulder |
| Claudia | Marin | Howard University |
| Charlie | Messina | University of Florida |
| Aditi | Misra | University of Colorado - Denver |

| Claire | Monteleoni | University of Colorado - Boulder |
|--------|-----------|----------------------------------|
| Melanie | Moses | University of New Mexico |
| Raj | Pandya | American Geophysical Union |
| Ann | Schwartz | Computing Community Consortium |
| Shashi | Shekhar | University of Minnesota |
| Jiayang | Sun | George Mason University |

Figure B-3. In-Person Workshop Attendee list. Participants in bold gave short presentations at the beginning of our breakout discussions to help frame and focus our conversations.

**Figure B-4. List of Virtual Workshop Attendees**

| Virtual Workshop Attendees | | |
|----------------------------|--|--|
| **First Name** | **Last Name** | **Affiliation** |
| Mara | Alagic | Wichita State University |
| Ali | Alghamdi | King Saud University |
| Moussa | Ali Abdou | Wascal |
| Chid | Apte | IBM Research |
| Sujata | Banerjee | VMware |
| M. Mehdi | Bateni | International Union of Soil Sciences |
| Rachel | Bellamy | IBM |
| Nadya | Bliss | Arizona State University |
| Kit | Boone | The University of Memphis |
| Zourkalaini | Boubakar | |
| Salem | Boumediene | University of Illinois - Springfield |
| Salma | Boumediene | Naval Postgraduate School |
| Liz | Bradley | University of Colorado - Boulder |
| Matthew | Burke | Amazon Web Services |
| Randal | Burns | Johns Hopkins University |
| Matt | Campbell | Oregon State University |
| Zuri | Chavers | |
| Chin-Wei | Chen | University of Washington - Seattle |
| Aurali | Dade | National Science Foundation |
| Richard | Donovan | University of California - Irvine |
| Sean | Downey | Ohio State University |
| Ram | Durairajan | University of Oregon |
| Behzad | Esmaeili | Purdue University |
| Adriane | Fernandes Minori | Carnegie Mellon University |
| Shannon | Fitzsimmons-Doolan | Texas A&M University Corpus Christi |
| Trent | Ford | University of Illinois |
| Johannes | Friedrich | World Resources Institute |
| Annarita | Giani | GE Research |
| Catherine | Gill | Computing Research Association |
| Sharon | Gillett | Microsoft Research |
| **Jared** | **Goldman** | Charles River Analytics |
| TG | Goodael | |
| Roger | Grant | FbSI |
| Haley | Griffin | Computing Research Association |
| Greg | Hager | Johns Hopkins University |

| Youngjib | Ham | Texas A&M University |
| Peter | Harsha | Computing Research Association |
| Thomas | Hauser | NCAR |
| Robin | Hoard | Hoard & Support.CO |
| **Seneca** | **Holland** | Texas A&M University Corpus Christi |
| Kathleen | Holman | Bureau of Reclamation |
| Pengyu | Hong | Brandeis University |
| Raya | Horesh | IBM Research |
| Yu | Hou | Western New England University |
| Sudhir | Jain | |
| Vandana | Janeja | University of Maryland Baltimore County |
| Anne | Johansen | National Science Foundation |
| Nima | Kargah-Ostadi | Callentis Consulting Group |
| John | Kemeny | University of Arizona |
| Deborah | Khider | University of Southern California |
| Rabinder | Koul | Vega MX Inc |
| Chandra | Krintz | University of California - Santa Barbara |
| **Yana** | **Kucheva** | The City College of New York |
| Piotr | Kulczakowicz | Quantum Startup Foundry, University of Maryland |
| Michaela | Labriole | New York Hall of Science |
| Sophia | Laettner | Jacksonville Global Shapers |
| Justin | Lancaster | Hydrojoule LLC |
| Yongcheol | Lee | Louisiana State University |
| Guann Pyng | Li | University of California Irvine - Calit2 |
| Gino | Lim | University of Houston |
| Asiyah | Lin | National Institute of Health |
| Dan | Lopresti | Lehigh University and CCC |
| Feng | Luo | Clemson University |
| Christine | Lv | University of Colorado - Boulder |
| W John | MacMullen | University of Illinois |
| Candace | Major | National Science Foundation |
| Matthew | McGoffin | University of California - Berkeley |
| Amy | McGovern | University of Oklahoma |
| Yohan | Min | Dartmouth College |
| Aditi | Misra | University of Colorado - Denver |
| Claire | Monteleoni | University of Colorado - Boulder |
| Melanie | Moses | University of New Mexico |
| Ashley | Mueller | USDA NIFA |
| Philip | Murphy | InfoHarvest Inc. |
| Gwen | Nero | Columbia University |
| Karen | Olcott | T-Mobile |
| Andrew | Padilla | Datacequia LLC |
| Juan | Padilla | Social Solutions, LLC |
| Indrani | Pal | Columbia University / CUNY |
| Raj | Pandya | American Geophysical Union |
| **Avi** | **Pfeffer** | Charles River Analytics |
| Kristian | Poe | University of California - San Diego |
| Yi | Qiang | University of South Florida |
| Lauren | Quigley | IBM Research |
| Yuhan (Douglas) | Rao | North Carolina State University |
| **Jon** | **Rask** | NASA Ames Research Center |
| Glen | Romine | NCAR |
| Rayan | Sadeldin | Schmidt Futures |
| Abolfazl | Safikhani | George Mason University |
| Sumeet | Sandhu | Climate Data Hub |
| Manikandan | Sathiyanarayan | National Taiwan University |
| Johannes | Schmude | IBM |
| Ann | Schwartz | Computing Research Association |
| Noelle | Selin | Massachusetts Institute of Technology |
| Ram | Shetty | Opex Systems LLC |
| Farahnaz | Soleimani | Oregon State University |
| Jing | Song | Genesis Codes Inc. |
| Jiayang | Sun | George Mason University |
| Tara | Tasuji | Technology and Information Policy Institute at the University of Texas at Austin |
| Hailay Zeray | Tedla | Addis Ababa University |
| Mukul | Tewari | IBM |
| Theo | Theoharis | Agoge Ventures |
| Lloyd | Treinish | IBM Thomas J. Watson Research Center |
| Ardhendu | Tripathy | Missouri University of Science & Technology |
| Charles | Wang | University of Florida |
| Jun | Wang | University of Iowa |
| Weichao | Wang | University of North Carolina - Charlotte |
| Wenwen | Wang | University of Georgia |
| David | Watkins | Michigan Technological University |
| Quarshie | Wordu | Kwame Nkrumah University of Science and Technology |
| Kay | Worthington | Australian National Research Lab |
| Heather | Wright | Computing Research Association |
| Helen | Wright | Computing Research Association |
| Chen | Xia | Penn State University |
| Yang | Xiao | University of Kentucky |
| Jinding | Xing | Carnegie Mellon University |
| K. | Xiong | University of South Florida |
| Jie | Xu | George Mason University |
| Shouhuai | Xu | University of Colorado - Colorado Springs |
| **Ellie** | **Young** | Common Action |
| Kai | Zhang | University at Albany |
| Peter | Zhang | Carnegie Mellon University |
| Weihang | Zhu | University of Houston |
| Woody | Zhu | Carnegie Mellon University |
| Zhigang | Zhu | The CUNY City College |

Figure B-4. Virtual Workshop Attendee list. Participants in bold gave short presentations after our breakout discussions to summarize the conversations held in each breakout session.